\documentclass[aps, prl, reprint]{revtex4-1}
\usepackage{amsmath}
\usepackage{graphicx}
\usepackage{float}
\usepackage{eqnarray}

\begin{document}

\title{Topology protection-unprotection transition: an example from multi-terminal superconducting nanostructures}
\date{\today}
\author{Xiao-Li Huang  }
\affiliation{Kavli Institute of NanoScience, Delft University of Technology, Lorentzweg 1, NL 2628 CJ, Delft, The Netherlands}
\author{Yuli V. Nazarov}
\affiliation{Kavli Institute of NanoScience, Delft University of Technology, Lorentzweg 1, NL 2628 CJ, Delft, The Netherlands}

\begin{abstract}
We show theoretically that in the superconducting nanostructures the gapped states of different topology are not always protected by separating gapless states. Depending on the structure design parameters, they can be either protected or not, with a protection-unprotection transition separating these two distinct situations. We build up a general theoretical description of the transition vicinity in the spirit of Landau theory. We speculate that similar protection-unprotection transitions may also occur for other realizations of topological protection in condensed matter systems. 
\end{abstract}

\maketitle

The topological ideas have been a source of inspiration in condensed matter for
many decades \cite{ref2}. In the last decade, there is an outburst of the experimental and theoretical activities related
to the topological materials and their unusual transport properties \cite{ref3-0,ref3-1,ref3-2}. For a material, the topology arises from and is determined by its bandstructure.
One of the remarkable results of the field is that there necessarily exist gapless states at the interface between the two insulators of different topology. Such gapless states are said to be {\it topologically protected}. The protected edge and surface states provide the transport
signatures of topology that are readily accessible for experimental research \cite{ref4-0,ref4-1,ref4-2}.


Many applications and realizations of various topological ideas in condensed matter physics are related to hybrid superconducting heterostuctures. Zero-energy Majorana states have been predicted \cite{ref5-0,ref5-1} and realized \cite{ref5-2,ref5-3} in such structures and remain in focus of active research. 
The Weyl points  in the spectrum of Andreev bound states of a four-terminal structure\cite{Weyl} have been predicted along with their robust transport signature of quantized transconductance \cite{transconductance} and associated spin effects \cite{WeylSpin}  A structure combining the topologies of three different kinds has been considered in \cite{OrderDisorder}.

A close analogue of topological isolators has been predicted and experimentally investigated in \cite{Omega}. The authors have studied the superconductivity induced in a normal-metal piece connected to three superconducting terminals. Typically, one expects a proximity gap to develop in the normal metal. It turns out that several  topologically distinct gapped phases can occur in the structure, those can be characterized by two integer topological numbers related to the number of windings of the semiclassical Green's function \cite{Omega,OrderDisorder,Amundsen2017}.  
These distinct phases are realized in different regions of the parameter space spanned by two superconducting phase differences between the terminals. General concept of topological protection implies that these regions are separated by finite strips of {\it gappless} phase. This has been probed by transport in the extra tunnel junction between the structure and a normal-metal lead. \cite{Omega}

We have found that such topological protection is not a universal property of a multi-terminal superconducting junction. Depending on parameters characterizing the junction desing, the protection may cease so that the different gapped phases are {\it not }
separated by a gapless strip. We show that a {\it protection-unprotection } 2nd order transition (PUT) separating these situation is described by a special Landau action that, in distinction from a common Landau action, gives rise to {\it two} order parameters above and below the transition - the gap and density of states at zero energy. We speculate that the proposed action is universal describing similar PUT's in general topological 
gapped phases.

\begin{figure}[H]
\includegraphics[width=0.8\linewidth]{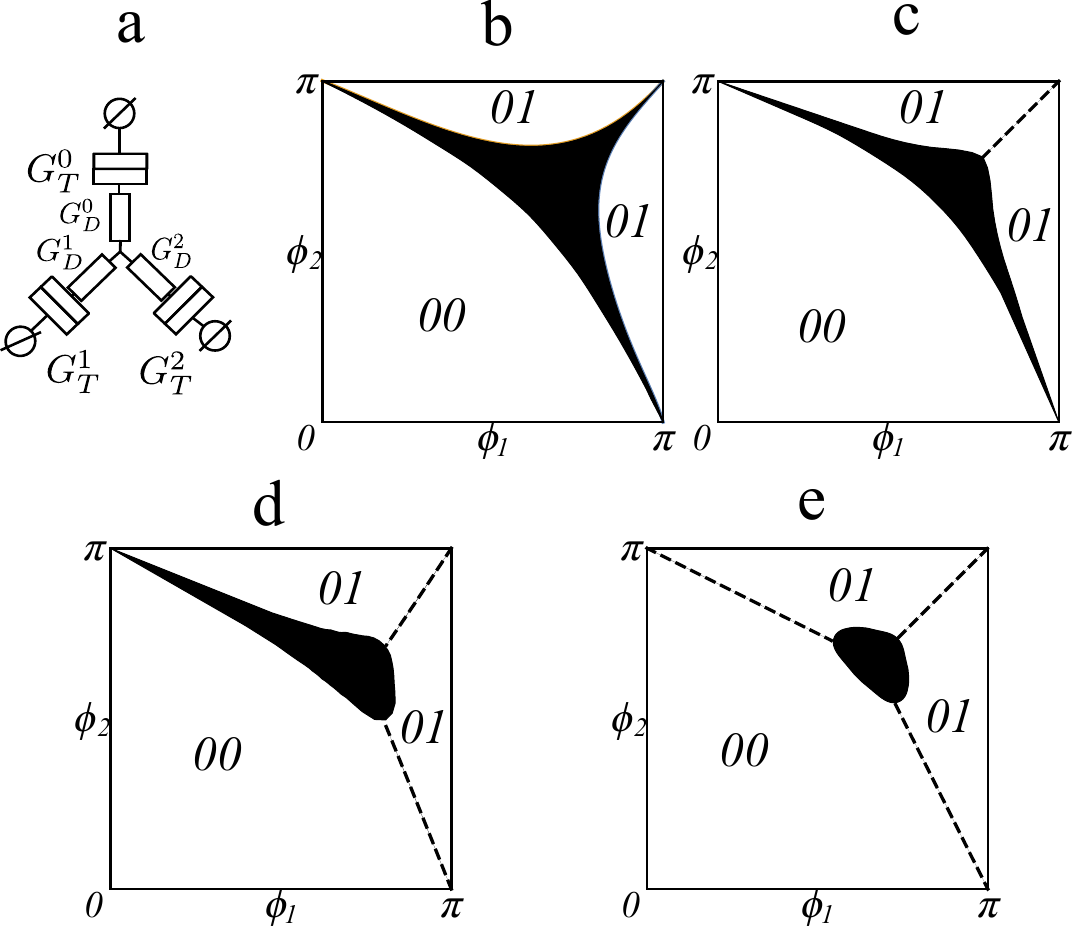}
\caption{Protection-unprotection transitions exemplified with a 3-terminal circuit (a). The protection of topologically distinct gapped phases $(N_{01},N_{10})$ implies the separation of their domains in  $\phi_1$-$\phi_2$ by the gapless state (black in (b-e)). The domain of the gapless state is thin near the special points $(0,\pi),(\pi,0),(\pi,\pi)$. Dashed lines indicate unprotected domain boundaries. The protection holds for all three points for sufficiently small diffusive conductances as compared to tunnel ones. Upon increasing the conductances, the protection ceases at special points with three protection-unprotection transitions. For calculations, all $G_T$ are taken the same and  $G^0_{D}=G^1_{D}=G^2_{D}=1.5 G_T$  (b);  $G^0_{D}=10, G^1_{D}=G^2_{D}=1.5$ (c); $G^0_{D}=G^1_{D}=10, G^2_{D}=1.5$(d); $G^0_{D}=G^1_{D}=G^2_{D}=10$ (e)}
\label{fig1}
\end{figure}

\ref{fig1}
Let us exemplify the PUT's with a three-terminal junction that is similar to the experimental system \cite{Omega}. The junction comprises three tunnel junctions adjacent to the terminals that are connected by diffusive pieces (Fig.\ref{fig1}a). 
The design parameters  are the conductances $G_Т^{i},G_D^{i}$, $i = 0,1,2$. We solve for the Green function in the structure at zero energy with the method described below. For the details of this derivation as well as for all other derivations, we refer to the supplementary material \cite{supplementary} .Generally, the Green function is a $2 \times 2$ Nambu matrix $g_z \sigma_z + g_x \sigma_х + g_y  \sigma_y$ parametrized by a unit vector $\vec{g}$, $\vec{g}^2=1$ where $g_z$ gives the density of states (d.o.s) at zero energy. Thus in a gapped phase, the vector is confined to $xy$ plane, and $g_{x,y} = \sin\mu, \cos\mu$, $\mu$ at a terminal equals to the corresponding superconducting phase $\phi$. The topological numbers are defined as \cite{Omega}
\begin{equation}
2 \pi N_{ij} =  \oint d {\bf r} \cdot \boldsymbol{\nabla} \mu + \phi_i - \phi_j
\end{equation}
where the integration contour goes from the terminal $i$ to the terminal $j$ and eventual jumps of $\mu$ along the contour are added upon projection of a jump to $(-\pi,\pi)$ interval \cite{Omega}. There are two independent topological numbers $n_{01},n_{02}$ and two independent phases $\phi_{1,2}$. In Fig.\ref{fig1} we plot the occurrence regions of each topological gapped phase in $\phi_1-\phi_2$ plane. Black gives the occurrence region of the gapless phase. In a fully protected situation (Fig.\ref{fig1}b) where $G_D$ are small or comparable with $G_T$, the topological phases are separated from each other by the gapless phase. The width of the separating region vanishes precisely at {\it special} points, where all phase differences are either $0$ or $\pi$. Upon decreasing $G_D$ we observe that the topological protection ceases step by step: the region of the gapless phase gets torn off one (Fig.\ref{fig1}c), two (Fig.\ref{fig1}d), and three (Fig.\ref{fig1}e) special points. The tearing off a point corresponds to a PUT. In an unprotected situation, the topological phases are separated by dashed lines where the gap is finite and the phase drop at a tunnel junction equals $\pm \pi$. In fact, similar PUT's have been seen in numerical simulations for a less realistic junction models \cite{Padurariu} but have received neither attention nor theoretical explanation.

In this paper, we would like to access the general situation: how a PUT occurs for a general $N$-terminal junction structure? We treat a general circuit design by means of quantum circuit theory where the junction is subdivided into nodes and connectors \cite{QuantumTransport}.
The Green functions in the nodes are obtained from the minimization of the action
\begin{equation}
{\cal S}= \sum_{c} S_c(\vec{g}_{c1} \cdot \vec{g}_{c2} )
\end{equation}
Here, the summation is over the connectors, $c1,c2$ denote the ends of a connector, that can be either nodes or terminals.
We resort to imaginary energy ($\varepsilon$) description where the vector $\vec{g}$ is conveniently real at any energy. The Green functions in N terminals are fixed to $\vec{g} = (\epsilon,\sin\phi,\cos \phi)/\sqrt{1+\epsilon^2}$, $\epsilon \equiv \varepsilon/\Delta$ being the energy in units of the superconducting energy gap $\Delta$ in the terminals.
In $N-1$-dimensional space of independent phases, we find $2^{N-1}-1$ special points where  
the phase differences are either 0 or $\pi$ and where the PUT's may occur.

Let us consider a common and rather degenerate case of $N=2$ where there is only a single superconducting phase difference $\phi$, the  special point is at $\phi=\pm \pi$ and the gapless phase may exist in this point only. The energy levels in the gap at this point correspond to transmission eigenvalues $T_n$ \cite{Beenakker}, $E_n = \Delta \sqrt{1-T_n}$. The protection corresponds to a transmission distribution that speads till $T \to1$, while unprotection corresponds to a distribution that ends at some $T_c <1$. The PUT corresponds to a "localization" transition reported in \cite{NazarovOld} that for a diffusive-tunnel structure takes place at $G_T=G_D$. 

Before concentrating on a PUT, let us understand the vicinity of a special point deep in protected regime. Let us note that at the special point and at zero energy all the nodes are separated in two groups located at $\vec{g} = (0,\pm 1,0)$ Owing to this, the solution for Green functions in this point is degenerate with respect to rotation about $y$-axis by angle $\psi$ and in each node $i$ can be parametrized as $\vec{g}_i = (\sin \theta_i \cos\psi, \cos \theta_i, \sin \theta_i\sin \psi)$. The deviations from the vicinity of the special point both in energy and the phases of the terminals lift the degeneracy and can be casted into the action of the following form
\begin{equation}
\label{eq:deepprotected}
{\cal S}/G = - \chi \cos \psi - \epsilon \sin \psi - r^2 \sin^2\psi
\end{equation}
Here, $G$ is a coefficient of the order of dimensionless junction conductance that is irrelevant for the minimization. Two topological phases corresponding to $\psi=0,\pi$ are realized in two $N-1$-dimensional regions that touch each other in the special point. There is a main axis orthogonal to the region surfaces in the special point, and $\chi$ stands for the deviation from the special point in the direction of this main axis. The $r$ gives the distance from the special point in all other $N-2$ directions perpendicular to the main axis. The term with ${\epsilon}$ pulls the Green functions in $z$-direction. With this, we can find the d.o.s in the nodes of the structure. While the maximum d.o.s. is node-specific, $\nu_{m,i} = \nu_0 \sin\theta_i$, its overall behavior is the same for all nodes,
\begin{equation}
\label{eq:dos_deepprotected}
\nu_i/\nu_{m,i} = \sqrt{1-(\chi/2r^2)^2}
\end{equation}
The d.o.s may be regarded as an order parameter in the gapless phase restricted by $|\chi|<\chi_c = 2r^2$ . Somewhat surprisingly, 
the action (\ref{eq:deepprotected}) can be also used to find the gap, a {\it complementary order parameter} for the gapped phase. Since the gap edge corresponds to a singularity in the Green function, the gap is found from the conditions of the action minimum and bifurcation,  $\partial S/\partial \psi = \partial^2 S/\partial \psi^2 =0$ that is satisfied at  imaginary $\epsilon= i \epsilon_g$,
\begin{equation}
\epsilon_g = \chi_c \left(\left(\chi/\chi_c\right)^{2/3}-1\right)^{3/2}.
\end{equation}
We illustrate the profiles of the d.o.s and the gap in Fig. 2 c,d.
\begin{figure}[h]
\includegraphics[width=.8\linewidth]{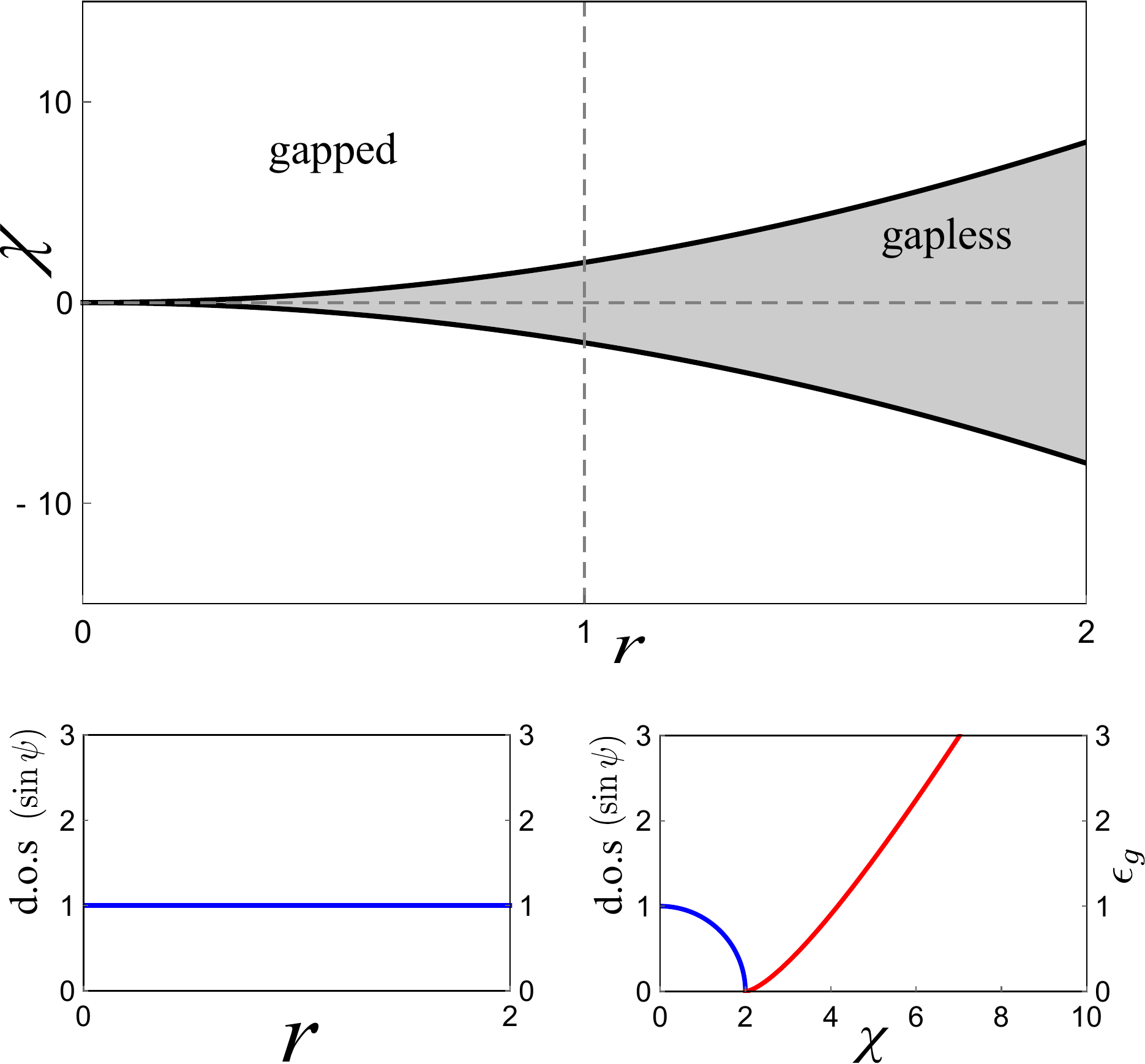}	
\caption{The vicinity of a special point in protected regime. Upper part: the domainds of gapless and gapped states the $\chi-r$ plane. Lower: the plots of the gap and d.o.s along the horizontal (left) and the vertical (right) dashed lines in the upper figure.}
\end{figure}

\begin{figure}[h]
\includegraphics[width=.8\linewidth]{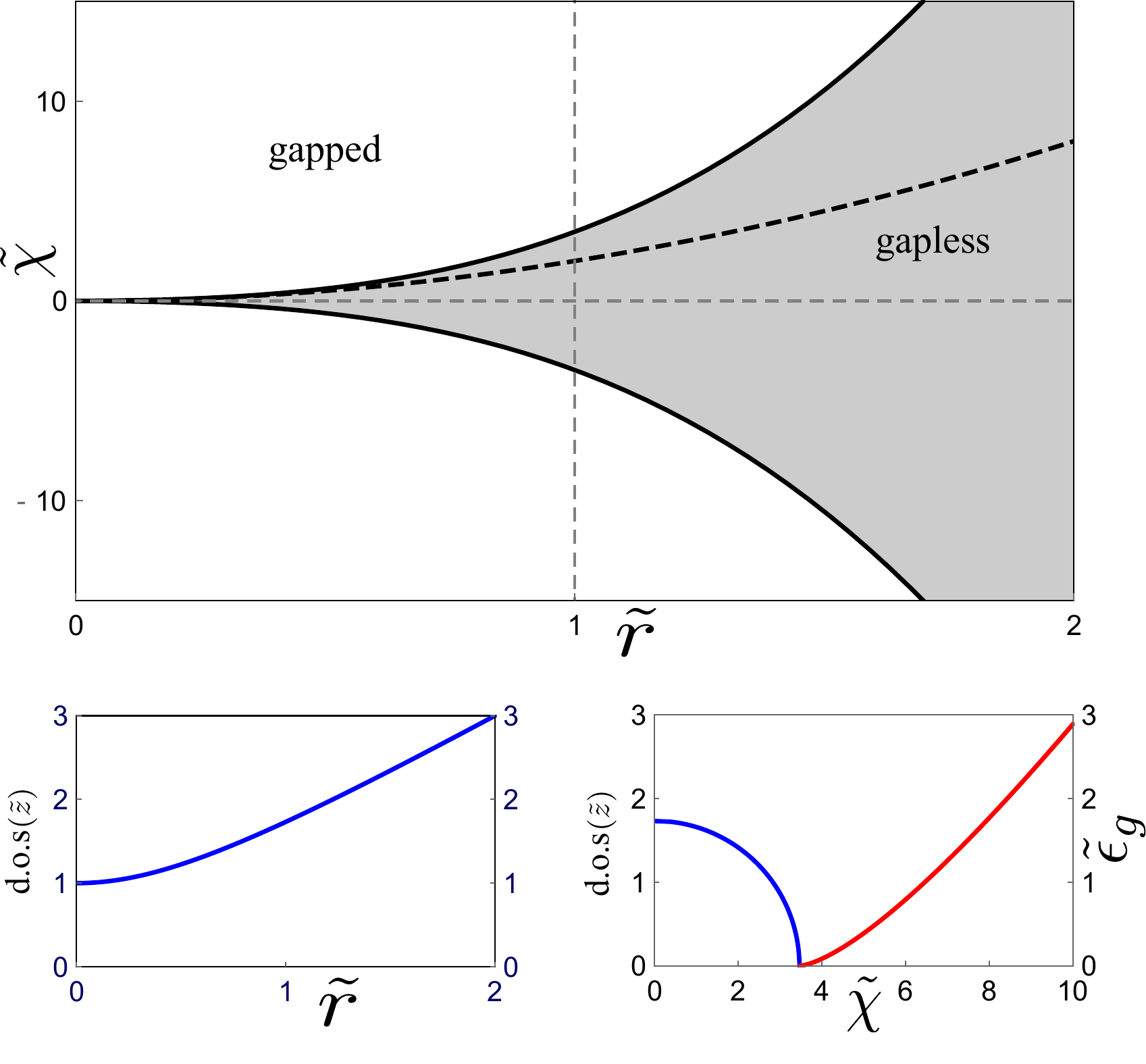}
\caption{The situation in the vicinity of a PUT, from the protected side of the transition. We use rescaled variables as defined in Eq. \ref{eq:rescaled}. Upper: the domains of gapped/gapless states in the $\tilde{\chi}-\tilde{r}$ plane. Dashed curve gives the asympotics of the domain boundary in deep protection regime $\tilde{\chi},\tilde{r} \ll 1$. Lower: the plots of the gap and d.o.s along the horizontal (left) and the vertical (right) dashed lines in the upper figure. }	
\end{figure}

Let us turn to the PUT description. Near a PUT, the Green functions nodes of a general structure are all close 
to one of the points $(0,\pm 1, 0)$. These two groups of nodes are connected by one or several connectors at which the phase drop is almost $\pi$.
The action can be expanded in a quadratic form with respect to the deviations of the Green functions $x_i,z_i$ from this point. If all eigenvalues of this quadratic form are positive,
the minimum is achieved at $x_i,z_i =0$ corresponding to unprotected situation. If at least one of the eigenvalues is negative, the action minimum is achieved at non-zero $x_i,z_i$ signaling formation of the gapless phase and topological protection. PUT corresponds to an eigenvalue crossing $0$. In spirit of Landau theory of the second-order phase transitions, we keep in the action the corresponding eigenmode only, $x,z$ being its deviations.

Taking into account the rotational symmetry at the special point, fourth-order terms and the anisotropies arising when the superconducting phases deviate from the point, we end up with the following action
\begin{equation}
\label{eq:main}
{\cal S}/G_L = \frac{a}{2}(x^2+z^2) +\frac{b}{4}(x^2+z^2)^4 - \chi x - \epsilon z - r^2 z^2
\end{equation}
Here, $a$ depends on the junction design and is the critical parameter that is negative for protected situation and zero at the PUT. In distinction from a common Landau action, the action 
(\ref{eq:main}) defines two complementary order parameters for gapless and gapped phases. The d.o.s in the gapless phase is proportional to $z$ at $\epsilon=0$ determined from the action minimization, while the determination of gap $\epsilon_g$ requires the extra bifurcation condition $\partial_{xx}{\cal S}\partial_{zz} {\cal S} - (\partial_{xz} {\cal S})^2 =0$. 

As usual, the action can be rescaled to convenient variables at a given value of $a$, either positive or negative,
\begin{equation}
\label{eq:rescaled}
{\cal S}/G_L = \frac{a^2}{b} \left(\pm\frac{\tilde{x}^2+\tilde{z}^2}{2} +\frac{(\tilde{x}^2+\tilde{z}^2)^4}{4} - \tilde{\chi} \tilde{x} - \tilde{\epsilon} \tilde{z} - \tilde{r}^2 \tilde{z}^2\right)
\end{equation}
where $\tilde{x}, \tilde{z} = x, z \cdot \sqrt{b/a}$, $\bar{\chi}, \epsilon= \sqrt{b/a^3} \cdot \, \tilde{\chi}, \tilde{\epsilon}$ and $\tilde{r}=r / \sqrt{a}$. 
In Figures 3, 4 we illustrate the profiles of 
the d.o.s. and gap in unprotected and protected regimes close to the PUT, respectively. We make use of the rescaled variables. We see that in the unprotected regime the distinct topological phases touch each other at $\chi=0$ and $\tilde{r} < 1/\sqrt{2}$.
The separating gapless phase always persists in the protected regime. At $\tilde{\chi},\tilde{r} \ll 1$ the action is reduced to that in deep protection regime (Eq.\ref{eq:deepprotected} )
giving $\tilde{\chi}_c = 2 \tilde{r}^2$. This asymptotics is plotted in Fig. 4 a with the dashed curve. 

Simple analytical formulas are obtained for the boundary between the gapped and gapless states, $\chi_c = 2\tilde{r}^2 \sqrt{2\tilde{r}^2 \pm 1}$, and for the gap at $\chi=0$, 
\begin{equation}
\epsilon_g = \frac{2}{3^{\frac{3}{2}}}(1-2r^2)^{\frac{3}{2}}
\end{equation}

\begin{figure}[h]
\includegraphics[width=.8\linewidth]{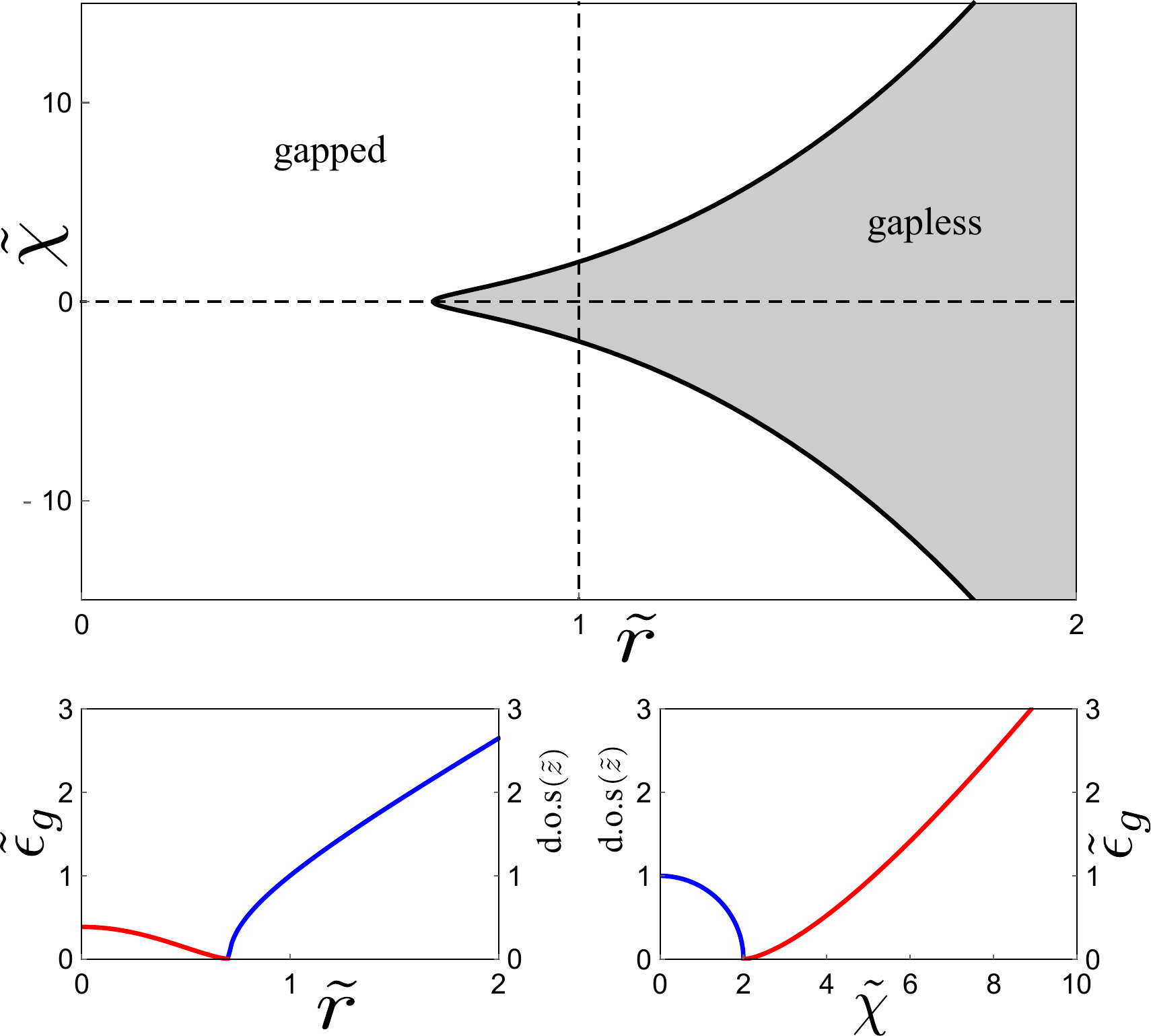}	
\caption{The situation in the vicinity of a PUT, from the unprotected side of the transition. We use rescaled variables as defined in Eq. \ref{eq:rescaled}. Upper: the domains of gapped/gapless states in the $\tilde{\chi}-\tilde{r}$ plane. The un $\tilde{\chi},\tilde{r} \ll 1$. Lower: the plots of the gap and d.o.s along the horizontal (left) and the vertical (right) dashed lines in the upper figure. }	
\end{figure}

In conclusion, we have studied the topological projection of distinct gapped states in $N$-terminal superconducting junction. The protection is manifested as a gapless state separating the gapped states in the parameter space of $N-1$ superconducting phases. We reveal that the protection may cease near special points as a result of a protection-unprotection transition in the parameter space of the junction designs.
We have found a Landau action that describes the the vicinity of the transiton. In distinction from common Landau actions, this one permits evaluation of complementary order parameters --- d.o.s. and gap --- for gapless and gapped states.


We speculate that the known generality of Landau actions would permit to extend our approach to a wider variety of topological phenomena in condensed matter physics. Such phenomena may include the gapping of the edge modes at the interfaces of distinct topological insulators that separate gapped states in real rather than parametric space. While the concrete physical mechanisms responsible for such PUT's may be involved and unknown, the essential phenomelogy of the transition may be captured by a Landau action in a form proposed in this Letter.

This work is part of the research programme of the Foundation for Fundamental Research on Matter (FOM), which is part of the Netherlands Organisation for Scientific Research (NWO). The authors acknowledge useful discussions with A. Akhmerov.

\bibliography{transtop}{}

\end{document}